\documentclass[prb,twocolumn,superscriptaddress,showpacs]{revtex4-1} 

\usepackage{graphicx}
\usepackage{times}
\usepackage{graphics}
\usepackage{epsfig}
\usepackage{epstopdf}
\usepackage{graphicx}
\usepackage{amsmath}

\begin{document}

\title{Multiphase density functional theory parameterization of the Gupta potential for silver and gold}

\author{John T. Titantah}
\affiliation{Department of Applied Mathematics, The University of Western Ontario,
 1151 Richmond St. N., London, Ontario, Canada, N6A\,5B7}
\email{jtitantah@gmail.com} 

\author{Mikko Karttunen}
\affiliation{Department of Chemistry, University of Waterloo,
200 University Avenue West,  Waterloo, Ontario, Canada, N2L\,3G1}
\email{mikko.karttunen@uwaterloo.ca}

\date{\today}




\begin{abstract}

The ground state energies of Ag and Au in the face-centered cubic (FCC), body-centered cubic (BCC), 
simple cubic (SC) and the hypothetical diamond-like phase, and  dimer were calculated as 
a function of bond length using density functional theory (DFT). These 
energies were then used to parameterize the many-body Gupta potential for Ag 
and Au. This parameterization over several phases of Ag and Au was performed to guarantee 
transferability of the potentials and to make them appropriate for studies of 
related nanostructures.  Depending on the structure, the energetics of the
surface atoms play a crucial role in determining the details of the nanostructure. 
The accuracy of the parameters was tested by performing a 2\,ns MD simulation of a cluster 
of 55 Ag atoms --  a well studied cluster of Ag, the most stable structure being 
the icosahedral one. Within this time scale, the initial FCC lattice was found to 
transform to the icosahedral structure at room temperature. The new set of parameters 
for Ag was then used in a temperature dependent  atom-by-atom deposition of Ag 
nanoclusters of up to 1000 atoms. We find a deposition temperature of 500$\pm$50\,K 
where low  energy clusters are generated, suggesting an optimal annealing temperature 
of 500\,K for Ag cluster synthesis. 
\end{abstract}

\maketitle

\section{Introduction}
Applications of noble metal nanoparticles are currently emerging in medicine where 
they are used as antimicrobial agents~\cite{kim07}, antivirals against HIV-1~\cite{lara11}, 
antiangiogenic agents~\cite{guranathan09}, in drug delivery~\cite{portney06} and in cancer 
therapy~\cite{portney06,huang06}. They are also used in chemical sensory devices due to their  
enhanced surface chemical activity~\cite{chen09,filippo09,he10,nezhad10,ngece11}.  Ag clusters 
as small as Ag$_7$ and Ag$_8$ supported with mercaptosuccinic acid have been demonstrated to 
be very potent in water purification through their chemical sensitivity to the presence of heavy 
metals like Pb, Cd, Hg~\cite{bootharaju11}. The application of noble metal nanoparticles in 
the electronic industry in inkjet printing of conductive lines for circuitory~\cite{lee05a} and as 
electronically conductive adhesives~\cite{chendapeng09,hsu07,lee05b,hu10} results from 
their  electronic properties. Organic memory devices based on DNA biopolymer nanocomposites 
with Ag nanoparticles have been demonstrated~\cite{hung11}. The large optical forces induced 
by the transfer of momentum from electromagnetic radiation to a dielectric nanoparticle also 
make them useful in nano optical manipulation ~\cite{ashkin70,ashkin86,liu10,filippo09}.
It has been shown that subjecting Au 
nanoparticles to optical forces induces aggregation and enhances 
the Raman scattering intensity of the thiophenol coverage of the Au nanoparticle~\cite{svedberg06}. 

It is common to  synthesize noble metal nanoparticle either as free standing or in an inert 
gas matrix with sizes ranging from one to several hundred of nanometers. Ag nanoclusters as 
small as 4.1 nm and 5.6 nm have been deposited on silicon substrates. Their thermally induced 
disintegration has been studied, revealing melting at temperatures well below Ag melting 
temperature~\cite{bhattacharyya09}. Clusters of few atoms are also routinely isolated by 
stabilizing them  with some protective molecules~\cite{lin09a,bhattacharyya09}. 
For example,  a 25 atom Au-thiolate cluster in solution~\cite{mcdonald11}, 
dodecanethiol-stabilized-Au$_{38}$~\cite{qian09}, a 1 nm lyzozyme-stabilized-Au nanocluster 
for Hg$^{2+}$ sensor~\cite{wei10} and DNA-encapsulated 10 atom Ag-cluster~\cite{petty10} 
have been reported.  A small angle X-ray study on the mechanism of Ag nanoparticle 
formation showed that nanoparticle formation initiates with the formation of Ag$_{13}$ 
clusters which agglomerate together to form a nanoparticle within 6 ms~\cite{takesue11}. 

The advances in synthesis methods have not been accompanied by an equal increase in 
understanding of the structural, electronic and optical properties. 
The nanosized nature gives these particles/clusters properties that are 
intermediate between molecular, which are of quantum origin, and bulk character. Size 
dependent structural changes have been widely investigated using methods ranging from 
experiments based on electron diffraction spectroscopy~\cite{reinhard97,blom06} to theoretical 
methods involving classical molecular mechanics approaches~\cite{xiaoli07} and quantum 
calculations. In particular, pseudopotential time-dependent density functional theory (TDDFT) calculations 
of the optical absorption of magic number noble metal nanoparticles have been carried out~\cite{weissker11}. 
Crossover from molecular to nanosized behavior was predicted to occur for clusters with 
more than 150 Ag atoms~\cite{weissker11}. Calculations of the absorption and the Raman 
enhancement of Ag$_n$-pyridine complexes (n=2-20) have been done using TDDFT and both 
properties were found to depend strongly on the cluster size~\cite{jensen07}.
 
Thoretical studies of noble metal clusters are widely based on classical approaches making 
use of interatomic potentials like the Gupta potential~\cite{cleri93,michaelian99}, the 
Sutton-Chen potential~\cite{sutton-chen,doye98} and the embedded atom potential~\cite{mei91,shibata02}. 
The latter has been used to study the size dependent melting of Ag 
nanocluster~\cite{zhao01}, spontaneous alloying in Au-Ag nanoclusters~\cite{shibata02} 
and structural optimization of Ag-Pd~\cite{wu11}. The Gupta potential is widely used in
predicting stable structures of noble metal nanoparticles like Ag and Au~\cite{pittaway09} 
and bimetallic clusters like Pd-Au~\cite{pittaway09}. The parameters of the potential 
are obtained by fitting experimental or DFT~\cite{pittaway09} data on the 
bulk face-centered cubic (FCC) system. This does not, however, guarantee transferability of the 
potential to the different phases of the system.  For example, when the resulting potentials 
are applied to lower coordination states of the material (such as  dimers, trimers, etc.), very 
short bonds and overbinding are obtained rendering the potential inappropriate for studies of  
low dimensional objects like nanoclusters. This effect has also been demonstrated to explain the 
finding that the Gupta and Sutton-Chen parameterizations predict  different growth patterns 
already for small Ag clusters~\cite{shao05}. Shao \textit{et al.}~\cite{shao05} pointed 
out that due to the flatter nature of the  dimer potential as given by the Gupta formulation, 
it is susceptible to yield more strain-tolerant structures than those generated using the 
Sutton-Chen potential.

In this study, we use the DFT approach to perform ground state energy calculations on the 
FCC, BCC, simple cubic, the hypothetical diamond-like phase and 
dimer. These energy profiles are used to parameterize the Gupta potential. The parameters 
are given as functions of the coordination number of the Ag and Au atoms-giving them 
a bond-order character. The appropriateness of the parameters for low coordinated 
structures is demonstrated.

\section{All electron calculation: parameterization of the Gupta potential}

We used the WIEN2k all-electron-full-potential linearized-augmented-plane-wave 
DFT code~\cite{wien2k} to calculate the binding energies of Ag and Au in 
FCC, BCC, simple cubic, diamond-like phases, and dimers. This DFT approach 
partitions the unit cell into muffin-tin (MT) spheres centered on each atomic site 
and the interstitial region. Within the MT spheres, the Kohn-Sham functions are given 
as linear combinations of spherical harmonics weighted by radial functions. In 
the interstitials they are parameterized as plane-waves.
The generalized gradient approximation 
(GGA)~\cite{perdew96} for the exchange and correlation energy is adopted. The two parameters governing 
the accuracy of the calculation are the number of plane-waves in the Brillouin zone 
and the plane-wave vector {\bf k} cut-off parameter - $RKM$. This latter parameter 
is the product of the maximum plane wave vector in the interstitial region and the 
smallest muffin-tin radius in the system. For our calculations, the total number of 
$k$-points ranged from 1000 to 3000, depending on the system and $RKM$ value of 6 
were found to converge the binding energies to an accuracy of 3\,mRy per atom. 

\begin{figure}
\begin{center}
\includegraphics[width=\columnwidth,trim=70px 60px 35px 70px, clip=true]
{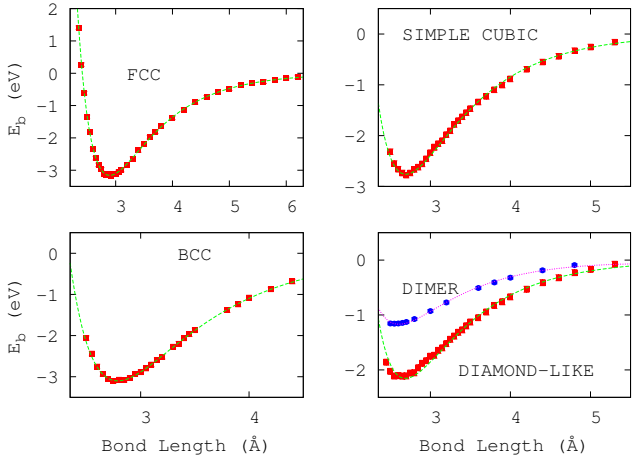}
  \caption{Energy per Ag atom as a function of distance between atoms for five phases of Ag: face-centered cubic, body-centerd cubic, simple cubic, Ag in the diamond-like phase and Ag dimer.}
\label{fig:E-r-ag}
\end{center}
\end{figure}

\begin{figure}
\begin{center}
\includegraphics[width=\columnwidth,trim=70px 60px 35px 75px, clip=true]
{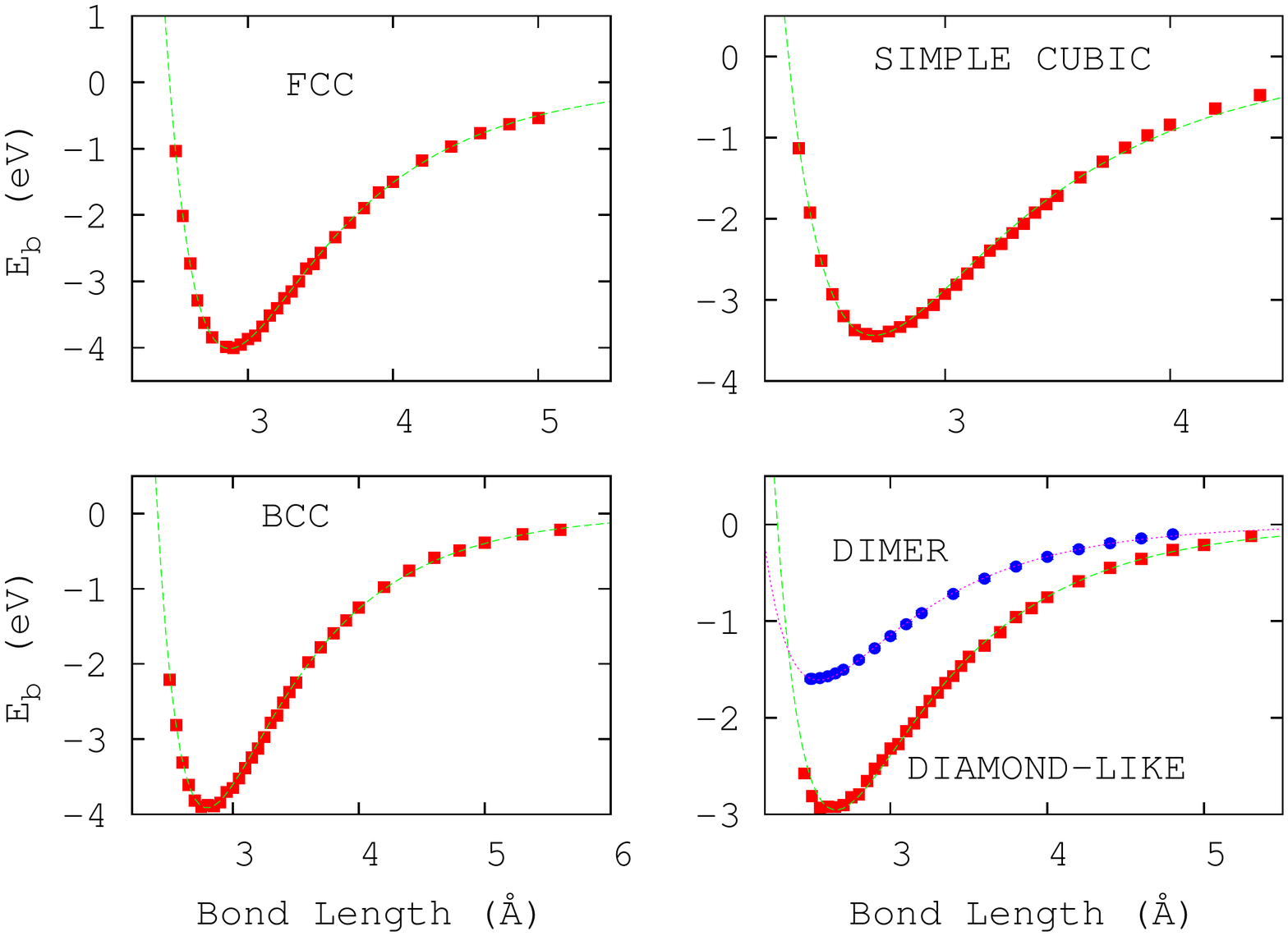}
  \caption{Energy per Au atom as a function of distance between atoms 
for five phases of Au: face-centered cubic, body-centerd cubic, 
simple cubic,  the diamond-like phase and Au dimer.}
\label{fig:E-r-au}
\end{center}
\end{figure}

\begin{figure}
\begin{center}
\includegraphics[width=\columnwidth,trim=80px 65px 40px 65px, clip=true]
{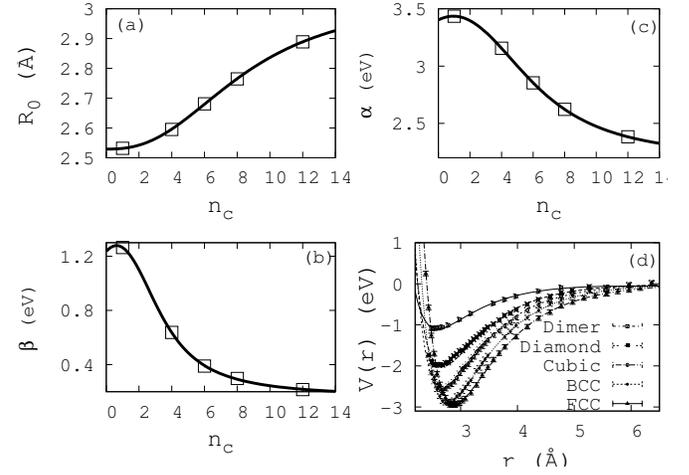}
  \caption{DFT-parameters of the Gupta potential for the four phases of Ag: face-centered cubic, BCC, simple cubic, Ag in the diamond-like phase and Ag dimer. The fitted potentials are shown in panel (d) and the points are the corresponding DFT binding energies.
}
\label{fig:params}
\end{center}
\end{figure}

Non-magnetic calculations were performed since both bulk Ag and Au are 
known to be non-magnetic. For each of the noble metal structures, energy-bond 
length dependencies were obtained. From them, extrapolation to very large lattice 
parameters permitted the isolated atom values to be removed yielding the binding 
energies as shown in Figs~\ref{fig:E-r-ag} and ~\ref{fig:E-r-au}.  For FCC Ag, 
we find an equilibrium lattice parameter of 4.07$\pm0.02$~\AA~and a binding 
energy of 3.16$\pm0.04$~eV/atom. The former compares very well with the measured 
value of 4.09~\AA~ while the binding energy overestimates the measured value of 
2.95 eV~\cite{kittel} by about 7~\%. The bulk modulus was found to be 106$\pm2$ GPa 
which is in excellent agreement with the experimental value of 109 GPa at 0\,K~\cite{neighbours58}.  
For Au,  a lattice parameter of 4.09$\pm0.02$~\AA~(see Fig.~\ref{fig:params}(a)) 
compares very well with the measured value of 4.08~\AA. The binding energy 
of 4.01$\pm$0.04 eV, however, overestimates  the experimental value of 3.81 eV~\cite{chantry12,needels92,kittel}. 
The bulk modulus of 179$\pm2$ GPa is in excellent agreement with experimentally measured value 
of 180 GPa at 0\,K~\cite{neighbours58}.

The BCC phase is less stable than the FCC phase by about 0.06 (0.08) eV/atom for Ag (Au) 
and is thus predicted to be the high pressure phase of these materials. This is in agreement 
with another theoretical work that predicted the same result for Au~\cite{sonderlind02}. 
The small energy difference between the BCC and the FCC phases of Ag is a possible 
reason for the recently observed continuous and reversible BCC-FCC phase transformation 
in Ag/V multilayers~\cite{wei11}. 

We used these energy-lattice parameter (or energy-bond length E-r) dependencies 
to parameterize the Gupta potential for Ag and Au. For this, we rescaled the 
binding energies by a factor of 0.934 (0.951) for Ag (Au) to equate the 
FCC value  with the measured value while rescaling distances by a factor of 
1.006 (0.996) for Ag (Au) to equate to the corresponding FCC experimental 
bond length of 2.889~\AA~(2.884~\AA). With these, the bulk moduli become 99 GPa 
for Ag and 170 GPa for Au. Most of the parameterizations that have been 
done are based on the bulk properties in the FCC 
phase~\cite{shao05, pittaway09}. 
To our knowledge, this is the first attempt to develop parameters that 
make the potential transferable enough for use in various phases. These new
parameters should therefore be appropriate in high temperature studies of these metals
and for low-dimensional structures formed from them. 

For this potential, the energy of atom $i$ is given by~\cite{clerirosato93}
\begin{equation}
V_i=V_i^r-V_i^a,
\label{eqn:vgup}\end{equation}
where the repulsive part of the potential is given by 
\begin{equation}
V_i^r={1 \over 2}\sum_{j\ne i}\alpha
\exp\left[-\lambda\left({r_{ij}\over R_0}-1\right)\right]
\label{eqn:repuls}
\end{equation}
and the attractive part by 
\begin{equation}
V_i^a={1 \over 2}\left[\sum_{j\ne i}\beta^2
\exp\left[-2\mu\left({r_{ij}\over R_0}-1\right)\right]\right]^{1/2}.
\label{eqn:attract}
\end{equation}

In the first approximation, the phase dependence of this potential 
is introduced by adopting a coordination dependence of the parameters; 
where coordination is defined by introducing a cut-off function $f_c$ 
\begin{eqnarray}
 f_c(x) &=& \Big{\{ }\begin{array}{cc}1& \mbox{ for $x\le3.1$~\AA} \nonumber\\
0&\mbox{ for $x>3.1~$\AA}\end{array}\nonumber\\
\label{eqn:cutoff}\end{eqnarray}
by which coordination $n_{ci}$ is given by
\begin{equation}
n_{ci}=\sum_{j\in \mathrm{Nei}(i)}f_c(r_{ij}),
\end{equation} where $\mathrm{Nei}(i)$ is the set of nearest neighbor atoms of atom $i$. 
The choice of a cut-off distance of 3.1~\AA~ was made to exclude 
the next nearest neighbor atoms of the BCC phase that show up beyond 3.2~\AA~ for both Ag and Au.

 The parameters $\lambda$ and $\mu$ are parameterized within the E-r dependence of FCC phases (Figs.~\ref{fig:E-r-ag} and ~\ref{fig:E-r-au}) as previous values for Ag~\cite{shao05} and Au~\cite{pittaway09} did not permit a good fit  of  our DFT calculations.
  
The parameters $\alpha(n_c)$, $\beta(n_c)$ and $R_0(n_c)$ are given  as  functions  of the coordination number of the atoms (and therefore of the phases). We suggest here the following simple functional forms:

\begin{eqnarray}
 \alpha(n_c) &=& \alpha_\infty\left(1+\zeta\L\left({ n_c+n_0\over \delta}\right)\right) \nonumber \\
  \beta(n_c) &=& \beta_\infty\left(1+\gamma \L\left({n_c+\Delta\over \eta}\right)\right)\nonumber \\
  R_0(n_c) &=&R_\infty\left[1-{\rho_0\over\left(1+(n_c/\nu)^\xi\right)}\right],
\label{eqn:params}\end{eqnarray}
where $\L(x)=2\left(\exp{\left(-x^2\right)}-1+x^2\right)/x^4$.

\begin{figure}
\begin{center}
\includegraphics[width=\columnwidth,trim=70px 60px 35px 65px, clip=true]
{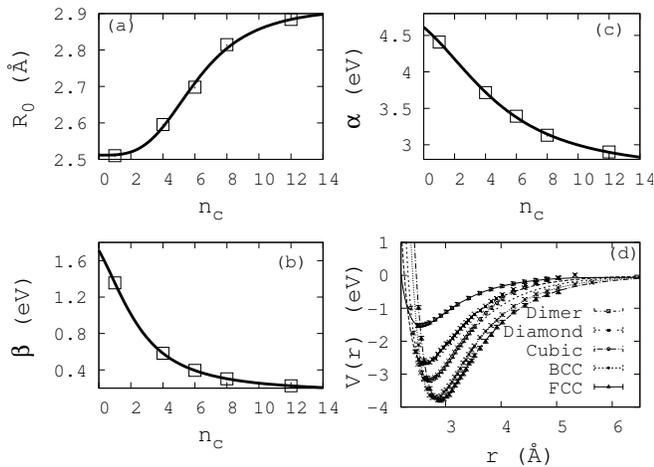}
  \caption{DFT-parameters of the Gupta potential for the five phases of Au: 
face-centered cubic, BCC, simple cubic, Au in the diamond-like phase and dimer. 
The fitted potentials are shown in panel (d) and the points are the corresponding DFT binding energies.
}
\label{fig:params-au}
\end{center}
\end{figure}

A fit of the parameters to these model functional forms are performed 
as shown in Figs.~\ref{fig:params} and ~\ref{fig:params-au} for Ag and Au, respectively. 
In the fitting procedure, periodic images are included such that interactions 
between atoms separated by distances as 
far as 9~\AA~are included. The parameters are listed in Table~\ref{tabl:params}. 

The effect of this new parameterization of Ag and Au clusters is that
in lower coordination environments (such as surfaces), the atoms become less 
energetic. For example, the energy  per atom of an Ag dimer was found to 
be 1.09 eV. This compares better with the experimental value of 0.83$\pm$0.02 eV ~\cite{morse86}
than the previously obtained values of 1.22~eV~\cite{shao05} and 1.40~\cite{wu11,negreiros10}. 
Also the dimer bond length of 2.59~\AA~is in good agreement with the measured 
value of 2.53~\AA~\cite{beutel93,wang04,morse86}~as compared to values of 2.37~\AA~\cite{shao05} and 2.43 ~\AA~\cite{wu11,negreiros10} that
result from the previous parameterizations. The best DFT dimer energy so far is 0.9\,eV with a 
corresponding dimer bond length of 2.58~\AA~\cite{nigam10}.
The Au dimer bond length of 2.56~\AA~ compares very well with the experimental 
value of 2.47~\AA~\cite{huber79} whereas previous parameterization sets this value 
at 2.3~\AA~\cite{pittaway09}.  The dissociation energy per atom of 1.53 eV for Au$_2$ is 
much closer to the experimental value of 1.15 eV ~\cite{james94} than the 
value of 2.42 eV that previous parameterization gives ~\cite{pittaway09}.
The observations made 
here about the properties of dimers are also valid for other low coordinations 
such as  2-, 3- and 4-coordinated atoms.

\begin{table*}[!ht]
\caption{Parameters of the Gupta potential for Ag and Au}
\begin{center}
\begin{tabular}{ccccccccccccccc}
 \hline
&$\lambda$&$\mu$&$\alpha_\infty$ (eV) &$\zeta$&n$_0$&$\delta$ &$\beta_\infty$(eV) &$\gamma$ &$\Delta$&$\eta$ &R$_\infty$(\AA) &$\rho_0$ &$\nu$&$\xi$\\
\hline
Ag&10.167&3.105&0.1610&6.946&-0.633&1.842&2.1633&0.588&-0.952&3.414&3.039&0.168&8.482&2.517\\
Au&12.696&3.179&0.1471&12.1218&1.268&1.989&2.5877&0.858&2.053&3.860&2.921&0.140 &6.114&3.436\\
 \hline
\end{tabular}
\end{center}
\label{tabl:params}

\end{table*}

The implication of the weaker binding of lower coordinated atoms --
as opposed to the strong binding obtained with the previous parameters~\cite{shao05,wu11,pittaway09} --
in studies of Ag and Au  nanoclusters is that surface atoms will become more mobile.  
This results in the nanoclusters experiencing surface premelting at lower temperatures. 
The current parameterizations may, thus, be very appropriate in surface studies.

\begin{figure}
\begin{center}
\includegraphics[width=\columnwidth,trim=70px 60px 35px 70px, clip=true]
{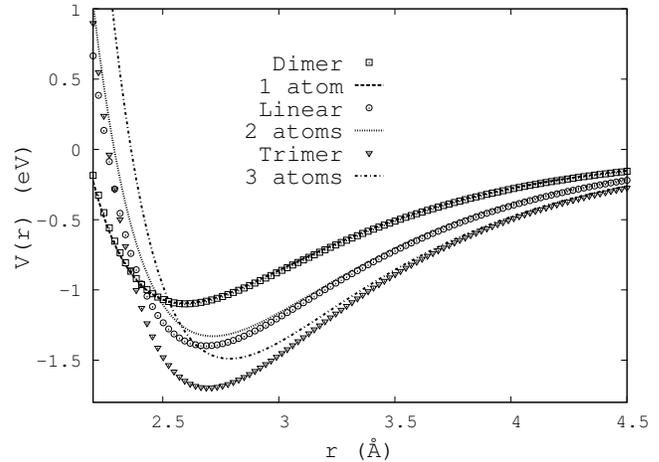}
\caption{The validity of the parameterization beyond r$_c$=3.1~\AA. 
Single atom, two atoms and three atoms are clearly seen to approach 
the central Ag atom in such a way that the potential of the 
central atom closely resembles that of an atom in a dimer, a linear 
chain and in a trigonal arrangement of Ag atoms, respectively.
}
\label{fig:approach}
\end{center}
\end{figure}

To check the validity of this parameterization for interactions beyond the 
cut-off distance of 3.1~{\AA} as defined in Eq.~\ref{eqn:cutoff}, we considered 
the case where $n_c=0$ and evaluated the parameters in Eq.~\ref{eqn:params}. 
The corresponding potential $U_0(r)$ was extracted from 
Eqs.~\ref{eqn:vgup} -~\ref{eqn:attract}. For this, the central atom is 
interacting with one, two or three equidistant atoms that are approaching 
it from beyond 3.1~\AA. In Fig.~\ref{fig:approach}, we plot this potential 
for one, two and three atoms approaching the central atom. We 
also show the potential of the atom in the case where it participates 
in a dimer, in a linear chain and in a trigonal arrangement. The 
single atom approaches with a potential that is very similar to that 
of the dimer; two atoms approach with a potential very close to that of 
the atom in a linear chain while three atoms approach like trimers. The 
potential described here thus correctly describes the short-, medium- 
and long-range interactions in Ag and Au. 

\section{Relaxation of Ag$_{55}$ cluster and MD deposition of Ag clusters}
\begin{figure}
\centering
\includegraphics[width=5cm,trim=60px 60px 35px 75px, clip=true]
{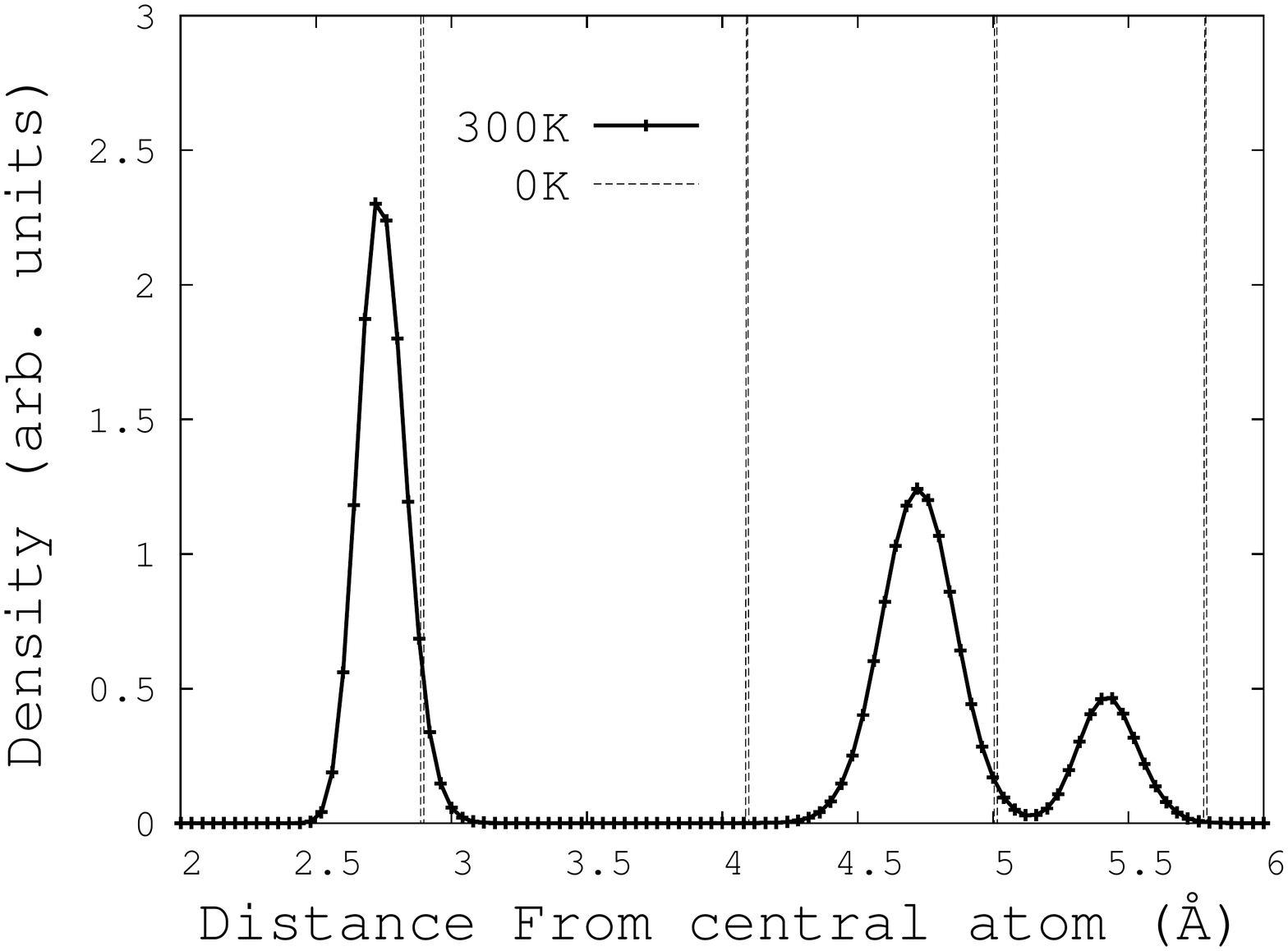}
\includegraphics[width=3.2cm,trim=10px 10px 10px 10px, clip=true]
{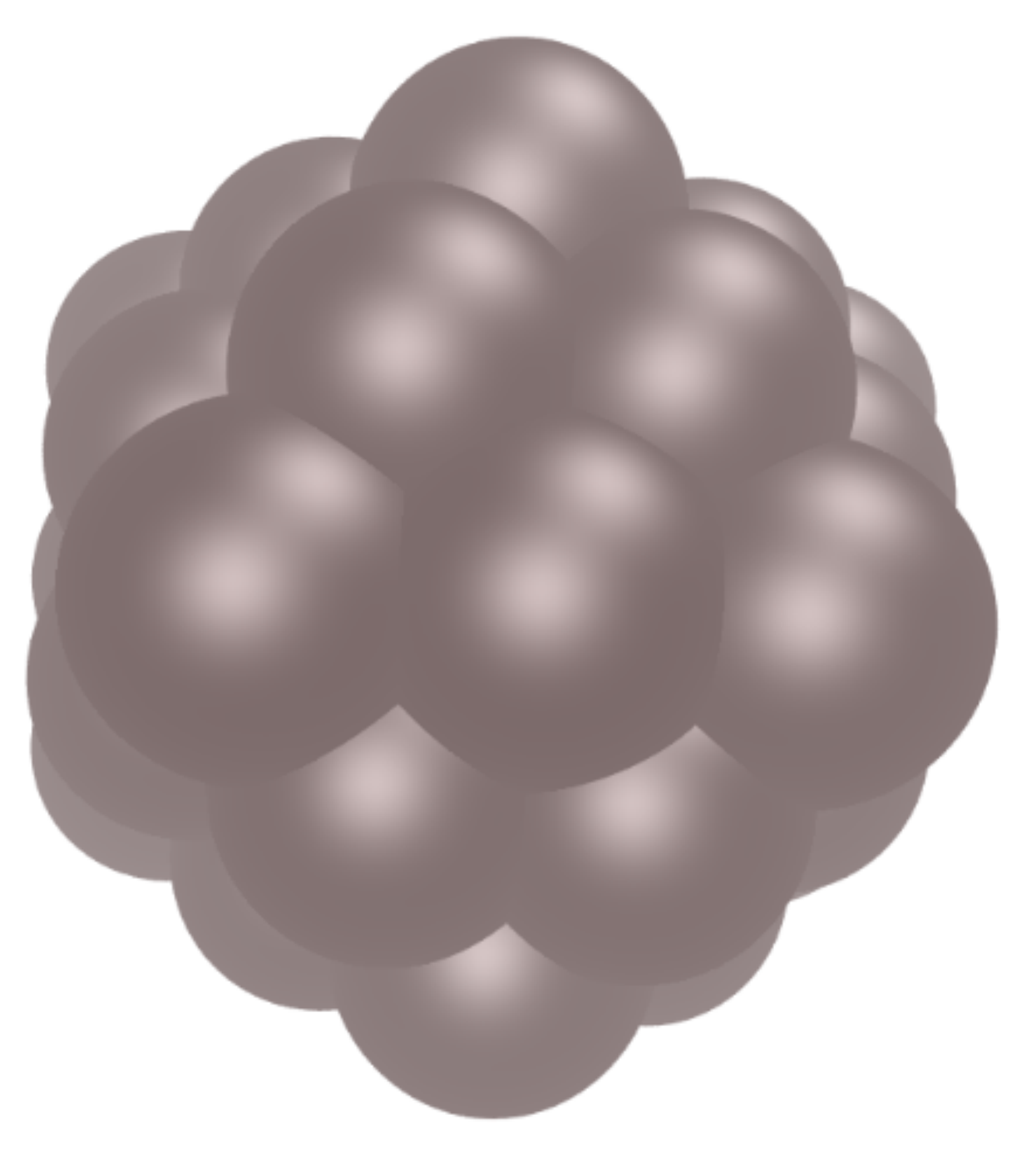}
\caption{Left: Density distribution as a function of distance from the central 
atom of a Ag$_{55}$ cluster for the unrelaxed (thin-dotted line) and the MD-relaxed structure 
at 300\,K (full line). 
The feature at 4.09~\AA~in the unrelaxed cluster 
shifts upwards and merges with the feature at 5.0~\AA~which shifts downwards.
 Right: Snapshot of MD-relaxed cluster at the end of the 2 ns simulation.
}
\label{fig:ag55}
\end{figure}

We performed a 2\,ns MD simulation to relax a 55 atom Ag cluster at 300\,K.  
Temperature was kept constant by using the Nos\'e-Hoover  
thermostat~\cite{nose-molphys84,hoover85}
and the velocity-Verlet 
scheme was adopted for the MD moves.  The cluster's structure was transformed from 
a 4-shell cluster to a 3-shell one within the first 10 ps.  Figure~\ref{fig:ag55} 
shows the density distribution as a function of distance from the central atom. 
The cluster is characterized by a central atom, a Ag$_{12}$ first-shell of 
radius 2.74~\AA, a next shell of 30 Ag atoms of radius 4.73~\AA~ and an outermost 
shell of 12 atoms of radius  5.45~\AA.  The structure of  the cluster at the end 
of the 2 ns is shown in the right of Fig.~\ref{fig:ag55}. 

To test further our parameters, we performed an atom by atom molecular dynamic 
deposition of Ag nanoclusters of up to 1000 atoms. The starting configuration 
was an Ag$_4$ cluster with four Ag atoms sitting at the corners of a square.  
While the cluster is undergoing relaxation at a fixed temperature, an atom is 
initiated far away (4 nm from the center of the formed cluster) with a kinetic 
energy of 2 eV and velocity pointing toward the cluster center, with the position 
of the atom chosen randomly on the sphere of radius 4 nm. Once this atom has 
entered the field of the relaxing cluster (distance less than 3.5~\AA~from 
the closest cluster atom) its dynamics are included in the Nos\'e-Hoover  
thermostating scheme. The resulting cluster was relaxed for 100 ps and the 
process was repeated for subsequent atoms.  A similar deposition process was 
done using the old parameters~\cite{shao05}. 

\begin{figure}
\begin{center}
\includegraphics[width=\columnwidth,trim=60px 60px 40px 60px, clip=true]
{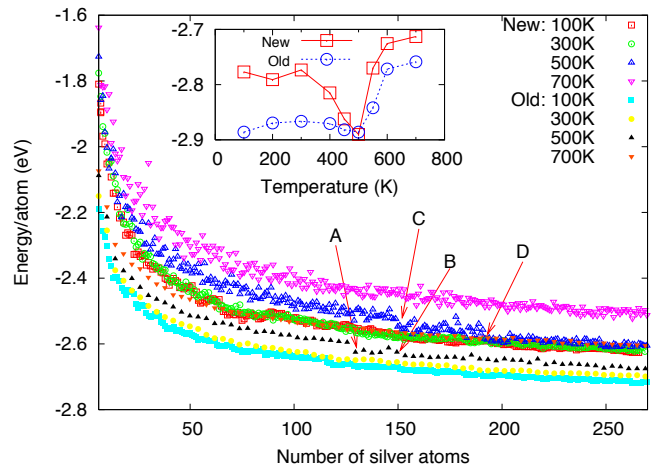}
\caption{The evolution of the energy per atom vs Ag cluster size 
at different deposition temperatures. Open symbols: new parameters,
closed symbols: old~\cite{shao05}.
 The arrows indicates points of energy drops. 
Inset: temperature dependence of $e_\infty$ (see Eqn.~\ref{eq:en}).}
\label{fig:ener-temp}
\end{center}
\end{figure}

Figure~\ref{fig:ener-temp} shows the evolution of the energy per atom as 
the cluster size grows at different temperatures using both the new 
and old sets of parameters.  We find that the energy decreases monotonically 
as the cluster size increases at all temperatures except at a temperature 
of 500$\pm$50\,K, where sudden drops in energy were recorded for cluster 
sizes larger than 126 and 150 for the old parameters, and 150 and 192 
for the new set of parameters. The cluster size of 150 is equal to the
 reported size beyond which a cross-over from molecular to nanometer 
optical behavior has been reported~\cite{weissker11}. We fitted the 
energy per atom as a function of the cluster size with the function ~\cite{huang11}
\begin{equation}
e(T)=E(N,T)/N=e_\infty(T)+bN^{-{1/3}}+cN^{-{2/3}},
\label{eq:en}\end{equation}
where $E(N,T)$ is the total energy of an N-atom Ag cluster at 
temperature T (in Kelvin).  The temperature dependence of the 
extrapolated large cluster energy, $e_\infty$, is shown in the inset 
and reveals lower energy clusters at a temperature of 500$\pm$50\,K. 
Note the big drop in energy as the temperature is raised from 300\,K 
to 500\,K for the new parameters and the almost constant value of 
the energy when the old parameters are used. This indicates that 
the former parameters may be capable of distinguishing between the 
various temperature induced structural changes within Ag nanoclusters. 
The finding of a more stable silver cluster at a temperature 
of 500$\pm$50\,K is in accord with temperature programmed Auger, 
low energy electron diffraction (LEED) and thermal desorption spectroscopy 
study of silver deposited on Mo(111) which demonstrated  that stable 
3-dimensional silver clusters are formed once treated at temperatures 
above 300\,K and below 650\,K~\cite{song01}. This temperature range is 
also in agreement with the finding of an Ag crystallization transition 
between 250 and 310 $^\circ$C where the agglomeration of silver atoms 
to form nanoclusters is favored~\cite{houk09}.

\begin{figure}
\begin{center}
\includegraphics[width=\columnwidth,trim=60px 65px 25px 50px, clip=true]
{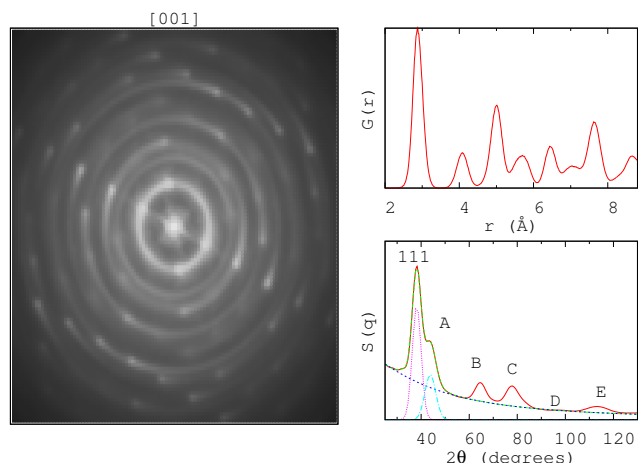}
  \caption{The lattice structure of the 561 atom Ag nanocluster generated 
by MD deposition at 500\,K and annealed at 300\,K. The electron diffraction 
pattern (left)  is calculated by using the Hartree-Fock parameterized form 
factors of Doyle and Turner~\cite{doyle}. The (111) spot can clearly be seen. Other spots  (200), (220), (311)/(222), (400) and (331)/420() are clearly visible and denoted as A, B, C, D and E, respectively. The density distribution with 
respect to the central atom, $G(r)$, (top-right) shows a highly ordered crystal 
and the structure factor $S(q)$ (lower-right) shows well defined features. 
The lines in the lower-right panel are for the functions used to fit the (111) spot. }
\label{fig:structure}
\end{center}
\end{figure}

Typical structural properties of a 561 atom Ag cluster atom-by-atom 
deposited at 500\,K and annealed at 300\,K with the new set of parameters 
is shown in Fig.~\ref{fig:structure}. We calculated the electron 
diffraction pattern by using the Hartree-Fock atomic form factors as 
parameterized by Doyle and Turner~\cite{doyle}. The atomic distribution 
with respect to the central atom of the cluster $G(r)$,  shows a highly 
ordered nanocluster, with FCC order up to a radius of more than 1\,nm. 
This is also confirmed by the structure factor $S(q)$,
 with $q={\sin\theta \over \lambda}$, $\lambda$ being the wave length 
and $\theta$ is the Bragg's diffraction angle. All results are averages 
over 10\,ns in time intervals of 200\,fs. The assignment of the spots agree 
excellently with that of Khan \textit{et al.} on Ag 
nanoparticles~\cite{khan11} and Kang ~\textit{et al.} on sintered 
inkjet-printed Ag nanoparticles~\cite{kang11}. In particular, 
the electron diffraction pattern is quite similar to the high resolution 
transmission electron microscopy (HRTEM) measurement of Ref.~\onlinecite{khan11} 
on Ag nanoparticles of size 16.37\,nm. $G(r)$ also shows that the 561 atoms 
cluster has a diameter of about 2.5\,nm, which is also confirmed by size $L$ 
of 24.97$\pm$0.21~\AA, obtained based on the position 2$\theta$=38.16$^\circ$ 
and the full-width at half-maximum (FWHM) $\sigma$=4.24$^\circ$ of the (111) 
spot as~\cite{klug54,khan11,govindaraj01}
\begin{equation}
L={0.94\lambda \over \sigma \cos2\theta},
\end{equation}where $\sigma$ is in radians and the wavelength $\lambda$=0.1541 nm 
is used. The 2$\theta$ and $\sigma$ values are obtained from a fit 
(in the interval $\pm$12$^\circ$ around the (111) spot) with a sum of 
three Gaussians and a Lorentzian, with the latter centered at $q$=0. 
A similar calculation on a 2093 atoms cluster cut out of bulk Ag 
and relaxed at 300\,K yielded a diameter of 4.22$\pm$0.29 nm which was 
also quite close to its size of 4.2 nm obtained on the basis of G(r).

\section{Conclusion}
We performed DFT calculations of the binding energies of Ag and 
Au in five different phases: FCC, BCC, simple cubic, diamond-like and 
dimer. We computed  the corresponding equilibrium bond lengths, binding energies 
and bulk moduli, and found them to compare very well with experimentally reported 
values.  We used the binding energy versus bond length dependencies 
to perform a cross-phase parameterization of the widely used many-body 
Gupta potential for Ag and Au. The new parameters, whose coordination-number 
dependencies are fitted to simple analytical functions, were found to correctly 
describe the energetic and structural behavior of low coordinated systems. 
We believe that these new parameterizations should be appropriate for studies 
of low dimensional structures such as nanoclusters 
and surfaces. 
They may also be used to reconcile the structural 
differences reported for small clusters of noble metals generated using 
the traditional bulk-fcc-based parameters of the Gupta potential -- 
which is susceptible to yield more strained structures -- and the Sutton-Chen potential, 
which leads to less-strained structures~\cite{shao05}. The current parameterization 
of the Gupta potential solves the problem of strain tolerance reported for the 
parameters of Shao \textit{et al.}~\cite{shao05} as the broadness at the bottom 
of the dimer potential well is improved.

\section*{Acknowledgements}
This work was supported by 
the Natural Sciences and Engineering Research Council of Canada (MK). 
Computational resources were provided by 
SharcNet [www.sharcnet.ca]. 



%

\end{document}